\documentclass[aip, 
jcp, longbibliography, 
twocolumn, floatfix, superscriptaddress, reprint, 10pt]{revtex4-1}
\usepackage[english]{babel}
\usepackage[utf8]{inputenc}
\usepackage[paperwidth=210mm,paperheight=297mm,centering,hmargin=2cm,vmargin=2.5cm]{geometry}
\usepackage[utf8]{inputenc}
\usepackage{verbatim}
\usepackage{bm}
\usepackage{amsmath}
\usepackage{amsfonts}
\usepackage{natbib}
\usepackage{pdfpages}
\usepackage{graphics}
\usepackage{float}
\usepackage{hyperref}
\usepackage[capitalize]{cleveref}
\usepackage{multirow}
\usepackage{booktabs}
\usepackage{makecell}
\usepackage{outlines}
\usepackage{lipsum}
\usepackage{siunitx}
\usepackage{mhchem}
\usepackage{caption}
\usepackage{subcaption}
\usepackage{xcolor}
\usepackage{ulem}
\usepackage{physics}
\usepackage{xparse}

\newcommand{\mbf}[1]{\ensuremath{\mathbf{#1}}}

\NewDocumentCommand{\rep}{s d<| d|>}{%
\IfBooleanTF{#1}{
   \IfValueTF{#2}{
       \IfValueTF{#3}{\braket{#2}{#3}}{\bra{#2}}
       }{
       \IfValueTF{#3}{\ket{#3}}{}
       }
   }{
   \IfValueTF{#2}{
       \IfValueTF{#3}{\braket*{#2}{#3}}{\bra*{#2}}
       }{
       \IfValueTF{#3}{\ket*{#3}}{}
       }
   }
}

\NewDocumentCommand{\rbra}{sm}{\IfBooleanTF{#1}{\rep*<#2|}{\rep<#2|}}
\NewDocumentCommand{\rket}{sm}{\IfBooleanTF{#1}{\rep*|#2>}{\rep|#2>}}
\NewDocumentCommand{\rbraket}{smom}{
    \IfBooleanTF{#1}{
        \IfNoValueTF{#3}{\rep*<#2||#4>}{\rep*<#2|#3\rep*|#4>}
    }{
        \IfNoValueTF{#3}{\rep<#2||#4>}{\rep<#2|#3\rep|#4>}
    }
}

\NewDocumentCommand{\field}{o m e{_} e{^} o e{_} e{^}}{
\IfValueTF{#5}{\overline{
  #2\IfValueT{#3}{_#3}\IfValueT{#4}{^{\otimes #4}} %
  \otimes
  #5\IfValueT{#6}{_#6}\IfValueT{#7}{^{\otimes #7}} %
  \IfValueT{#1}{;#1}
}}{
  \IfValueTF{#4}{\overline{
     #2\IfValueT{#3}{_#3}\IfValueT{#4}{^{\otimes #4}}
     \IfValueT{#1}{;#1}
  }}
  {#2\IfValueT{#3}{_#3}}
}
}

\NewDocumentCommand{\frho}{o e{_} e{^}}{
\field[#1]{\rho}_{#2}^{#3}
}

\newcommand{\e}{a}  %

\newcommand{\br}{\mbf{r}}
\newcommand{\bx}{\mbf{x}}

\newcommand{\brhat}{\hat{\mbf{r}}}
\newcommand{\faca}{c}
\newcommand{\facb}{d_n}

\NewDocumentCommand{\ex}{e_}{
\IfValueTF{#1}{\e_{#1}\bx_{#1}}{\e\bx}
}  %

\NewDocumentCommand{\lm}{e_}{
\IfValueTF{#1}{l_{#1}m_{#1}}{lm}
}
\NewDocumentCommand{\nlm}{e_}{
\IfValueTF{#1}{n_{#1}\lm_{#1}}{n\lm}
}
\NewDocumentCommand{\enlm}{e_}{
\IfValueTF{#1}{\e_{#1}\nlm_{#1}}{\e\nlm}
}
\NewDocumentCommand{\en}{e_}{
\IfValueTF{#1}{\e_{#1}n_{#1}}{\e n}
}
\NewDocumentCommand{\nlk}{e_}{
\IfValueTF{#1}{n_{#1}l_{#1}k_{#1}}{nlk}
}
\NewDocumentCommand{\enlk}{e_}{
\IfValueTF{#1}{\e_{#1}\nlk_{#1}}{\e\nlk}
}
\NewDocumentCommand{\enl}{e_}{
\IfValueTF{#1}{\en_{#1}l_#1}{\en l}
}

\NewDocumentCommand{\nnl}{s}{
\IfBooleanTF{#1}{n_1 n_2 l}{n_1; n_2; l}
}
\NewDocumentCommand{\ennl}{s}{
\IfBooleanTF{#1}{\en_1 \en_2 l}{\en_1; \en_2; l}
}

\NewDocumentCommand{\gslm}{s}{
\IfBooleanTF{#1}{\sigma\lambda\mu}{\sigma;\lambda\mu}
}

\newcommand{\Rhat}{{\hat{R}}}
\newcommand{\ihat}{{\hat{i}}}

\newcommand{\mc}[1]{{\color{blue}{ #1}}}
\newcommand{\MC}[1]{\mc{\bf MC: #1}}
\newcommand{\fm}[1]{{\color{purple}{ #1}}}
\newcommand{\gf}[1]{{\color{red}{ #1}}}

\newcommand{\CHF}[3]{{}_{1}F_{1}\left(#1 ,#2, #3 \right)}

\newcommand{\GA}[1]{\Gamma\left(#1\right)}
\newcommand{\Ylm}[1]{Y_l^m\left(#1\right)}
\newcommand{\lmax}[0]{{l_\mathrm{max}} }
\newcommand{\nmax}[0]{{n_\mathrm{max}} }

\newcommand{\nel}[0]{{n_\mathrm{species}} }
\newcommand{\nneigh}[0]{{n_\mathrm{neigh}} }
\newcommand{\nfeat}[0]{{n_\mathrm{feat}} }

\newcommand{\ntrain}[0]{{n_\mathrm{train}} }

\newcommand{\nactive}[0]{{n_\mathrm{active}} }
\newcommand{\fcut}[0]{{f_\mathrm{cut}} }
\newcommand{\rcut}[0]{{r_\mathrm{cut}} }

\newcommand{\D}[2][]{\ensuremath{\mathop{}\!\mathrm{d}^{#1}{#2}\,}}

\newcommand{\rij}{r_{ij}}

\newcommand{\denf}{g_{\sigma}}

\newcommand{\CC}{C\nolinebreak\hspace{-.05em}\raisebox{.4ex}{\tiny\bf +}\nolinebreak\hspace{-.10em}\raisebox{.4ex}{\tiny\bf +}}
\def\CC{{C\nolinebreak[4]\hspace{-.05em}\raisebox{.4ex}{\tiny\bf ++}}}

\newcommand{\maxnote}[1]{{\color{cyan} #1}}
\newcommand{\alexnote}[1]{{\color{teal} #1}}

\newcommand{\rascal}[0]{\texttt{librascal}}
\newcommand{\todorev}[1]{{}}
\makeatletter
\AtBeginDocument{\let\LS@rot\@undefined}
\makeatother
\begin{document}

\setcitestyle{super}

\title{Efficient implementation of atom-density representations}

\author{F\'elix Musil}
\thanks{These authors contributed equally to this work.\\Corresponding author: \texttt{felix.musil@epfl.ch}}
\author{Max Veit}
\thanks{These authors contributed equally to this work.\\Corresponding author: \texttt{max.veit@epfl.ch}}
\affiliation{Laboratory of Computational Science and Modeling, Institute of Materials, \'Ecole Polytechnique F\'ed\'erale de Lausanne, 1015 Lausanne, Switzerland}
\affiliation{National Center
for Computational Design and Discovery of Novel Materials (MARVEL)}

\author{Alexander Goscinski}
\author{Guillaume Fraux}
\author{Michael J. Willatt}
\affiliation{Laboratory of Computational Science and Modeling, Institute of Materials, \'Ecole Polytechnique F\'ed\'erale de Lausanne, 1015 Lausanne, Switzerland}

\author{Markus Stricker}
\affiliation{Laboratory for Multiscale Mechanics Modeling, Institute of Mechanical Engineering, \'{E}cole Polytechnique F\'{e}d\'{e}rale de Lausanne, 1015 Lausanne, Switzerland}
\affiliation{Interdisciplinary Centre for Advanced Materials Simulation,
Ruhr-University Bochum, Universit\"atsstra\ss{}e 150, 44801 Bochum, Germany}

\author{Till Junge}
\affiliation{Laboratory for Multiscale Mechanics Modeling, Institute of Mechanical Engineering, \'{E}cole Polytechnique F\'{e}d\'{e}rale de Lausanne, 1015 Lausanne, Switzerland}

\author{Michele Ceriotti}
\affiliation{Laboratory of Computational Science and Modeling, Institute of Materials, \'Ecole Polytechnique F\'ed\'erale de Lausanne, 1015 Lausanne, Switzerland}
\onecolumngrid
\begin{abstract}

Physically-motivated and mathematically robust atom-centred representations of molecular structures are key to the success of modern atomistic machine learning (ML) methods.  They lie at the foundation of a wide range of methods to predict the properties of both materials and molecules as well as to explore and visualize the chemical compound and configuration space.
Recently, it has become clear that many of the most effective representations share a fundamental formal connection: that they can all be expressed as a discretization of $n$-body correlation functions of the local atom density, suggesting the opportunity of standardizing and, more importantly, optimizing the calculation of such representations.
We present an implementation, named \rascal,  whose modular design lends itself both to developing refinements to the density-based formalism and to rapid prototyping for new developments of rotationally equivariant atomistic representations.
As an example, we discuss SOAP features, perhaps the most widely used member of this family of representations, to show how the expansion of the local density can be optimized for any choice of radial basis set. 
We discuss the representation in the context of a kernel ridge regression model, commonly used with SOAP features, and analyze how the computational effort scales for each of the individual steps of the calculation.
By applying data reduction techniques in feature space, we show how to further reduce the total computational cost by at up to a factor of 4 or 5 without affecting the model's symmetry properties and without significantly impacting its accuracy.

\end{abstract}
\twocolumngrid

\maketitle

\todorev{AUTHOR COMMENT KEY

\maxnote{Max Veit}

\fm{Félix Musil}

\alexnote{Alexander Goscinski}

\gf{Guillaume Fraux}

\mc{Michele Ceriotti}
}

\section{Introduction}

\begin{figure*}[t]
    \centering
    \includegraphics[width=0.9\textwidth]{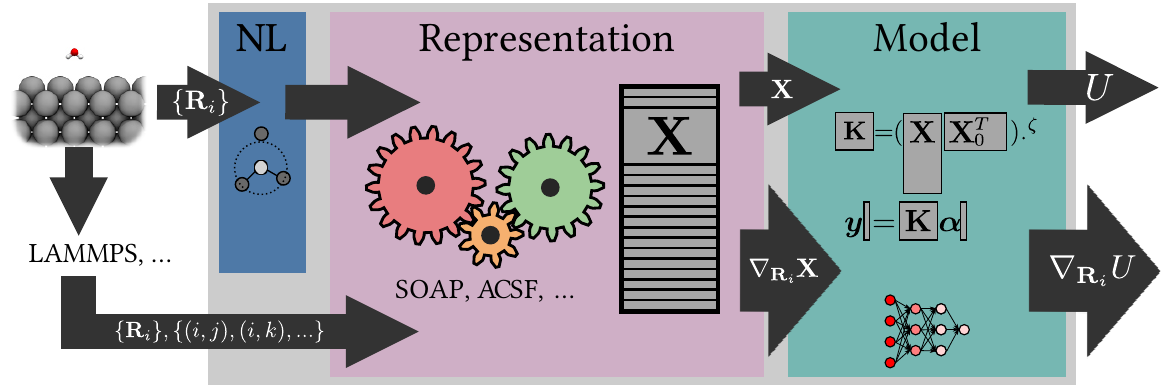}
    \caption{A scheme showing the different components involved in the evaluation of energies and forces for an atomic-scale machine-learning model. These steps need to be performed for each new structure in a screening procedure, or each step in a molecular-dynamics simulation.}
    \label{fig:librascal-schematic}
\end{figure*}

Supervised and unsupervised machine learning (ML) methods are gaining increasing importance in the field of atomistic materials modeling.
Supervised ML methods -- in particular, those used to construct interatomic potentials (MLIPs)\cite{Behler2007,bart+10prl,Shapeev2017,ferreLearningMolecular2017,hirnWaveletScattering2017,Schutt2017,Glielmo2018,drau19prb}, find structure-property mappings across chemical compound space\cite{Rupp2012QM7,Montavon2013,Bartok2017,musilMachineLearning2018}, and model the dependence on configuration and composition of experimentally relevant quantities like the dipole moment~\cite{gris+18prl,veitPredictingMolecular2020,Christensen2019,unke-meuw19jctc}, polarizability~\cite{wilkins2019}, band structure~\cite{chan+19npjcm,benmahmoudLearningElectronic2020}, and charge distribution~\cite{Ghasemi2015,alre+18cst,Grisafi2019} -- are useful tools in the quest for predictive materials modelling, specifically the use of large, complex, quantum-accurate simulations to access experimental length and time scales~\cite{sossoNeuralNetwork2012,Behler2017,Caro2018,Cheng2019,Mocanu2018,veit+19jctc,rosenbrockMachinelearnedInteratomic2019,deringerGeneralpurposeMachinelearning2020,deri+21nature}.
Furthermore, unsupervised ML methods are gaining prominence as a way to interpret simulations of ever-increasing complexity~\cite{rohrdanzDeterminationReaction2011,perez-hernandezIdentificationSlow2013,huanAcceleratedMaterials2015,gasparottoRecognizingLocal2018,ceriottiUnsupervisedMachine2019,rogalNeuralNetworkBasedPath2019,helfrechtStructurePropertyMaps2020}.

All of these methods fundamentally rely on a transformation of the system's atomic coordinates into a form amenable to the construction of efficient and transferable machine-learning models.
Usually, this implies that the features that represent an atomic configuration reflect the transformations (invariance or covariance) of the target properties with respect to fundamental symmetry operations, and that the prediction of the extensive properties of a structure is decomposed into that of local contributions, written as a function of a description of the neighborhood of individual atoms\cite{Behler2007,bart+10prl}.

We will focus, for the majority of this paper, on the problem of \emph{regression} of a property expressed as a function of these transformed coordinates (hereafter called just ``representation'').
By far the most common application of structure-property regression in the context of atomistic simulations is in the fitting of potential energy surfaces, which are used in molecular simulations or to compute thermodynamic averages.
The majority of the considerations we make here applies to the prediction of any scalar property of the system, although the calculation of gradients might be less important than in the case of potentials, and we use MLIP and ``model'' interchangeably in what follows.  
Figure~\ref{fig:librascal-schematic} schematically illustrates the typical procedure of a single timestep in an atomistic machine learning molecular dynamics (ML-MD) simulation.  After the atomic coordinates are read in, and the neighbor list is computed to determine the  local environments around each atom, the coordinates are transformed into an intermediate representation.  It is this representation that is then passed to the machine learning model -- be it one based on neural networks (NN)~\cite{Behler2011}, Gaussian process regression (GPR)\cite{rasmussenGaussianProcesses2006}, or one of several other closely-related methods.
The accuracy and the transferability of the regression model %
are usually greatly improved by the use of representations that fulfill the requirements of symmetry and locality,
\cite{Behler2011,soap-prb, chmi+18nc, Willatt2019}
while at the same time being sensitive to all relevant structural changes~\cite{soap-prb,onat+20jcp,parsaeifardAssessmentStructural2020}, being smooth functions of the atomic coordinates, and -- ideally -- being free of degeneracies which map completely different structures to the same descriptor~\cite{pozdnyakovCompletenessAtomic2020}.

Here we focus on a class of representations that fulfill these requirements, and that can be constructed starting from a description of a structure in terms of an atom density - which is naturally invariant to permutations of the atom labels - which is made translationally and rotationally invariant by first summing over $\mathbb{R}^3$, and then averaging $\nu$-points correlations of the resultant atom-centered density over the $O(3)$ improper rotation group~\cite{Willatt2019}.
The smooth overlap of atomic position (SOAP) power spectrum is perhaps the best-known member of this class of representations~\cite{soap-prb}, but a wealth of other descriptors such as those underlying the spectral neighbor analysis potentials (SNAP)~\cite{thompsonSpectralNeighbor2015}, the atomic cluster expansion (ACE)~\cite{drau19prb,bachmayrAtomicCluster2020}, moment tensor potentials (MTP)~\cite{Shapeev2017}, as well as the equivariant extensions  $\lambda$-SOAP\cite{gris+18prl} and the N-point contractions of equivariants (NICE)\cite{nigamRecursiveEvaluation2020} representation can be recovered as appropriate limits or extensions.  Atom-centred symmetry functions\cite{Behler2007,Behler2011} can also be seen as a projection of these atom-density representations on a bespoke set of basis functions.

However, even though these representations are related through a common mathematical formalism~\cite{Willatt2019,drau19prb}, the cost of evaluating them, and the accuracy of the resulting models, can vary greatly.
In some cases, different frameworks have been shown to yield comparable errors\cite{Nguyen2018}, while other studies have suggested a trade-off between accuracy and computational cost, with the combination of SOAP features and Gaussian process regression (hereafter termed just SOAP-GAP) emerging as the most accurate, but also the most computationally demanding method.\cite{zuoPerformanceCost2020,rosenbrockMachinelearnedInteratomic2019}

In fact, the evaluation of SOAP features and their gradients can take anywhere from \SIrange{10}{90}{\percent} (depending on the system and the parameters chosen) of the total computational cost of the energy and force evaluation in a typical molecular dynamics (MD) simulation with the SOAP-GAP method; almost all of the remaining cost is taken up by the evaluation of kernel (and its gradients) required to compute the GAP energy and forces. We therefore discuss optimization strategies aimed at reducing the computational cost of these two critical components.
While we focus, at present, on a serial implementation, the modular structure that we introduce to optimize single-core performance is also very well-suited to parallelization, which becomes indispensable when aiming at extending simulation size and timescale.

We begin in Section~\ref{sec:theory} with an overview of density-based representations, and present our benchmarking methodology in Section~\ref{sec:methods}. We continue in Section~\ref{sec:impl-bench}, showing how the mathematical formulation of density-based representations reveals several opportunities for optimization, which we implement and systematically benchmark.  We then expand these benchmarks to a variety of realistic simulation scenarios, shown in Section~\ref{sec:comparative_benchmarks}, comparing against an existing simulation code and investigating the effect of convergence parameters.  We present further experimental extensions of the code's capabilities in Section~\ref{sec:experimental_features}.  Finally, in Section~\ref{sec:conclusion}, we summarize the improvements and describing the role of our new, modular, efficient code \rascal{} in the modern atomistic ML ecosystem.

\section{Theory}
\label{sec:theory}
We begin by giving a brief overview of the construction of a symmetrized atom-density representation~\cite{Willatt2019}, introducing the notation we use in the rest of this paper to indicate the various components that are needed to evaluate the features associated with a given structure.
The construction operates by a sequence of integrals over symmetry operations, applied to a smooth (or Dirac-$\delta$-like) atom density that is taken to describe the structure. After summing over translations, one obtains a description of the atomic environment $A_i$
around the \emph{central atom} $i$, that depends on the neighbor positions $\br_{ji} = \br_j - \br_i$. Each atomic species $\e$ is associated with a separate density built as a superimposition of Gaussian functions $\denf(\bx)\equiv \rep<\bx||\denf>$ with variance $\sigma^2$, restricted to a local spherical cutoff $\rcut$ by a smooth function $\fcut$:
\begin{equation}
    \rbraket{\ex}{A; \rho_i} = \sum_{j\in A_i} \delta_{\e \e_j}
    \rep<\bx-\br_{ji}||\denf> \fcut(\rij).
    \label{eq:density-x}
\end{equation}
We use a notation that mimics the Dirac bra-ket formalism~\cite{willattFeatureOptimization2018}, in which the bra indicates the entity being represented (the density field $\rho$ centred on atom $i$ of structure $A$) and the ket the indices that label different features (in this case, the chemical species $\e$ and the position at which the field is evaluated, $\bx$, that serves as a continuous index).
To simplify the notation, when discussing the construction of the features for an arbitrary configuration we omit the reference to the atomic structure such that $\rbraket{\ex}{A; \rho_i}\rightarrow\rbraket{\ex}{\rho_i}$.
Following Ref.~\citenum{Willatt2019}, we introduce the symmetrized ($\nu+1$)-body correlation representation
\begin{multline}\label{eq:body-order-x}
\rep<\ex_1; \ldots \ex_{\nu}||\frho_i^{\nu}> \\
=\sum_{k=0,1}\int_{SO^3} \!\!\!\! \D{\Rhat}
\rep<\ex_1|\Rhat\ihat^k\rep|\frho_i> \ldots
\rep<\ex_{\nu}|\Rhat\ihat^k\rep|\frho_i>,
\end{multline}
where $\frho_i^{\nu}$ is a tensor product of $\nu$ atom centered fields averaged over all possible improper rotations.
This object can be understood as a fixed $\nu$-point stencil centered on atom $i$ which is applied continuously to the density field, hence accumulating correlations of body order $\nu+1$.
To perform the rotational average, it is convenient to expand the atom density on a basis whose expansion coefficients are given by
\begin{multline}
\!\!\!\rbraket{\enlm}{\rho_i} = \!\!\sum_{j\in A_i} \delta_{\e\e_j} \!\int\! \D{\bx}  \rep<nl||x>\! \rep<\lm||\hat{\bx}> \! \rep<\bx-\br_{ji}||\denf>,
    \label{eq:density-nlm}
\end{multline}
where $x=\norm{\bx}$, $\hat{\bx}=\bx/x$.
$\rbraket{x}{nl}\equiv R_{nl}(x)$ are orthogonal radial basis functions, which may or may not depend explicitly on $l$ (see e.g. Ref.~\citenum{Caro2019}), and  $\rbraket{lm}{\hat{\bx}}\equiv \Ylm{\hat{\bx}}$ are spherical harmonics.
As we discuss in Section~\ref{sub:radial_integral}, the choice of $\rbraket{x}{nl}$ is flexible and can thus be guided by considerations of computational and information efficiency. 
The angular dependence, on the other hand, is most naturally expanded using spherical harmonics, which results in compact expressions for the density correlation features of \cref{eq:body-order-x} in terms of products and contractions of the expansion coefficients of \cref{eq:density-nlm} (see Ref.~\citenum{Willatt2019} for more details).
For the case of $\nu=2$,
\begin{multline}
\rep<\ennl||\frho_i^2> = \frac{1}{\sqrt{2l+1}} \\
\sum_m (-1)^m \rep<\en_1 lm||\frho_i> \rep<\en_2 l(-m)||\frho_i>
\label{eq:nu2-basis-coupled}
\end{multline}
which corresponds to the SOAP power spectrum of Ref.~\citenum{soap-prb} up to some inconsequential factors.
In the following text we discuss and benchmark the efficient implementation of the density expansion and spherical invariant of order 2, i.e. the power spectrum, in \rascal.

\begin{figure*}[tbhp]
    \centering
    \includegraphics[width=0.8\textwidth]{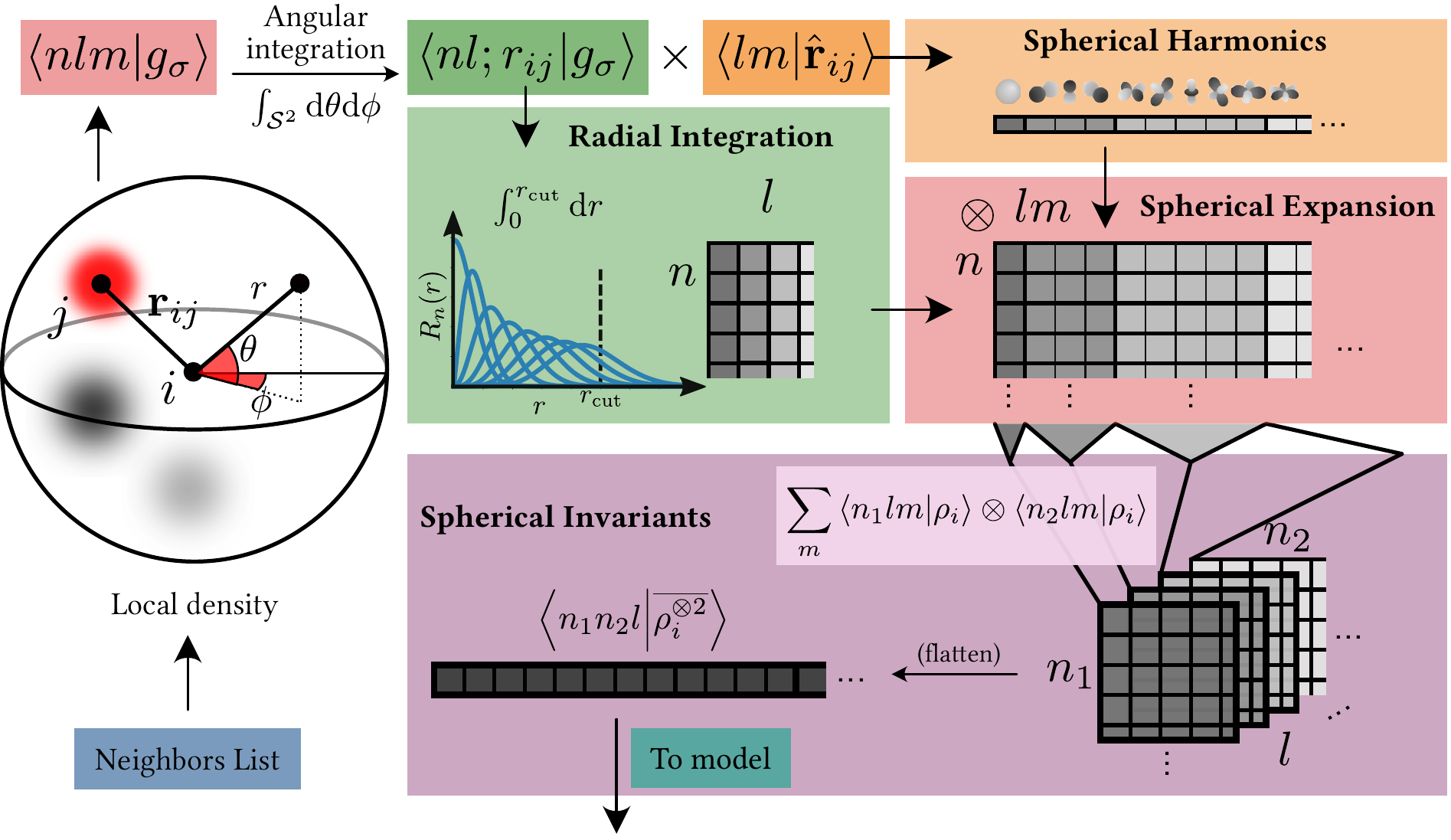}
    \caption{Schematic showing the process of expanding the density in a radial and angular basis set, and recombining those to form spherical invariants (or covariants).}
    \label{fig:spherical_expansion_schematic}
\end{figure*}

\section{Methods}
\label{sec:methods}

In order to provide a concrete assessment of the impact of the optimizations we describe in this paper, and to demonstrate the performance of the optimized code on a variety of realistic systems, we apply a comprehensive benchmarking strategy that compares different classes of systems and breaks down the overall computational cost into contributions associated with different steps of the evaluation of the ML model.
We focus on a typical use case of SOAP features, namely, as the input to a kernel ridge regression (KRR) model for a property and its derivatives (e.g. energy and forces). Even though \rascal{} focuses on the evaluation of features, including the model evaluation step is crucial to assess the computational effort of the evaluation of the features in the context of the overall cost of the model. 
In the same spirit of assessing the computational effort in a way that reflects the most common use case scenario, we focus on the cost of evaluating %
a previously-trained model, rather than the cost to train the model itself.
The training step is almost always limited by memory, not computation time, and must only be performed once per potential.  When running a simulation, such as an MD trajectory, the representation and model evaluation constitute the real limiting factors in what can be achieved with a given potential in terms of statistics, system size, and complexity of the target properties. %

We report and examine the benchmarks separating the logical components of the overall calculation, as summarized in Figure~\ref{fig:librascal-schematic} and~\ref{fig:spherical_expansion_schematic} -- namely the construction of the neighbour list, the calculation of the local density expansion (that can be further broken down into the evaluation of radial and angular terms) the combination of the density coefficients to obtain an invariant representation, and the evaluation of the model itself.  Most of these steps can also be broken down into the time required to compute just the representation (energy) versus the overhead for computing gradients (forces) in addition.

For the representation stage, it is possible to track the computational cost as a function of a few key parameters, namely the radial and angular expansion limits ($\nmax$ and $\lmax$, respectively).  The benchmarks for this stage are reported not per atom, but per \emph{pair}, consistent with the overall scaling of this component of the calculation.  The timings reported in this way are therefore also mostly independent of the system in question, i.e., the variation between systems is usually comparable to the variation between individual timing runs.
The model stage has less of a detailed dependence on the spherical expansion parameters, but the system dependence is more subtle.  The main influence on the computational cost is the feature space dimension $\nfeat$ and the number of environments $\nactive$ used to parametrize the model.  As will be discussed in Section~\ref{sub:dim-reduction-sparsification}, both of these parameters can be reduced significantly by the use of dimensionality reduction algorithms, with reduced computational cost generally trading off with the accuracy of predictions (a pattern seen in many other machine learning frameworks\cite{zuoPerformanceCost2020}).
In order to run and organize the large number of individual benchmarks required for this study, we have made extensive use of the signac data management framework\cite{signac_commat,signac_zenodo}, which can be accessed from an open repository\cite{rascal_benchmarks}.

\subsection{Datasets}
\label{sub:methods-datasets}

The system dependence of the overall computation is influenced by two major factors. The first is the number density, which -- together with the cutoff radius $r_\text{cut}$ -- determines the total number of pairs that must be iterated over to compute the representations, as well as the number of the degrees of freedom needed to fully characterize the local environment, which in turns affect the radial and angular expansion parameters necessary to represent it.
The second is the number of chemical elements that are present, which directly affects the dimensionality of the representation. However, several optimizations are possible depending on the model and species composition, as well as the distribution of these species throughout the system, making this a subtle and nontrivial influence on the total model cost.

Therefore, we have decided to benchmark the overall cost (neighbour list, representation, and model together) on a selection of five realistic datasets that represent both typical and challenging applications of machine learning potentials.  For a single-species system, we have chosen the bulk silicon dataset\cite{kermodeSiliconTesting2018} from \citet{Bartok2018}; despite its simple species composition, it still represents a large array of structural diversity.  The fluid methane dataset\cite{veitBulkMethane2018} from \citet{veit+19jctc} has two chemical species, but distributed homogeneously throughout the cell; the dataset additionally contains a range of different cell densities.  In order to include more challenging multi-species systems, we have selected three additional datasets from different application areas.  The solvation dataset from \citet{rossiSimulatingSolvation2020} consists of structures each containing a single molecule of methanesulfonic acid within a large cell of liquid phenol, where the presence of multiple species and the inhomogeneity of their distribution presents a challenge for both representation and fitting algorithms.  The molecular crystal dataset ``CSD1000r'' used in \citet{musi+19jctc} contains up to four species, where not all species are present in each separate structure.  Finally, the widely-used QM9 dataset\cite{rama+14sd} contains isolated molecules of up to 9 heavy (non-hydrogen) atoms each, and composed up to five chemical species -- where, again, not every species is represented in every structure.

\section{Implementation of Invariant Representations}
\label{sec:impl-bench}

We begin by discussing the \rascal{} implementation of the  the power spectrum SOAP features, and by showing how a deeper understanding of the structure of the atom-density correlation features can be exploited to improve substantially the cost of evaluation.
Benchmarks on all the datasets discussed above are included in the SI -- here we choose a subset of the different test cases, since in most cases the computational cost can be normalized in a way that minimizes the dependence on the specifics of the system at hand.

\subsection{Density coefficients}
\label{sub:radial_integral}

The exact expression for the density coefficients depends on the specifics of the atom density field and on the basis used to expand it.  To see this, it is advantageous to separate the integral in Eq.~\eqref{eq:density-nlm} into radial and angular coordinates.  The angular integral is most naturally expressed using the spherical harmonics basis, as evidenced e.g. by the particularly simple form of \cref{eq:body-order-x}.
Then, regardless of the choice of the functional form of the atom density or the radial basis, the density coefficients can be written as a sum over functions of neighbor distances and orientations
\begin{equation}\label{eq:density-coefficients}
\rep<\enlm||\frho_i> =
\sum_{j\in A_i} \delta_{\e\e_j}  \fcut(r_{ji}) \rep<\nlm;\br_{ji}||\denf>
\end{equation}
where
\begin{equation}
\rep<\nlm;\br||\denf> =
\!\int\! \D{\bx}  \rep<nl||x>\! \rep<\lm||\hat{\bx}> \! \rep<\bx-\br||\denf>.
\end{equation}
For a Gaussian atom density, the integral can be factorized into
\begin{equation}
\rep<\nlm;\br||\denf> = \rep<\brhat||lm> \rep<nl;r||\denf>,
\label{eq:spherical_expansion_coefficients_factorized}
\end{equation}
containing a radial integral
\begin{equation}
\!\!\!\!\rep<nl;r||\denf>= 4\pi e^{-\faca r^2}\!\! \int_0^\infty \!\!\!\dd{x} x^2 \rbraket{nl}{x} e^{-\faca x^2}\mathsf{i}_{l}\!\left(2\faca x r\right),
    \label{eq:radial-integral-gaussian-density}
\end{equation}
where $c=1/2\sigma^2$, and the radial and angular degrees of freedom are explicitly coupled by the $l$ dependence of the modified Bessel function.
Thus, the density coefficients can be computed by evaluating spherical harmonics and radial integral functions for each pair of neighbors, and then summing over their products
\begin{equation}
\!\!\!\rep<\enlm||\frho_i> = \!
\sum_{j\in A_i} \delta_{\e\e_j}  \fcut(r_{ji})\!
\rep<\brhat_{ji}||lm>\! \rep<nl;\rij||\denf>\label{eq:spherical_expansion_coefficients_gaussians}
\end{equation}
Alternative atom-centred density formulations such as in ACE\cite{drau19prb} or TurboSOAP\cite{Caro2019} lead to similar expressions for the radial function.
For instance, TurboSOAP chooses a Gaussian atomic density that is symmetric about $\br_i$ instead of $\br_{ji}$, making it possible to factorize the radial term such that $\rep<nl;\rij||\tilde{g}>=\rep<n||\rij>\rep<l||\rij>$.  Both terms can be efficiently computed using recurrence relations in $l$ and $n$.
In \rascal, the density expansion is implemented only for Gaussian atomic densities symmetric about $\br_{ji}$, using two types of radial basis sets: The Gaussian type orbital (GTO) basis and the discrete variable representation (DVR) basis.

\begin{figure}[bhtp]
    \centering
    \includegraphics[width=0.5\textwidth]{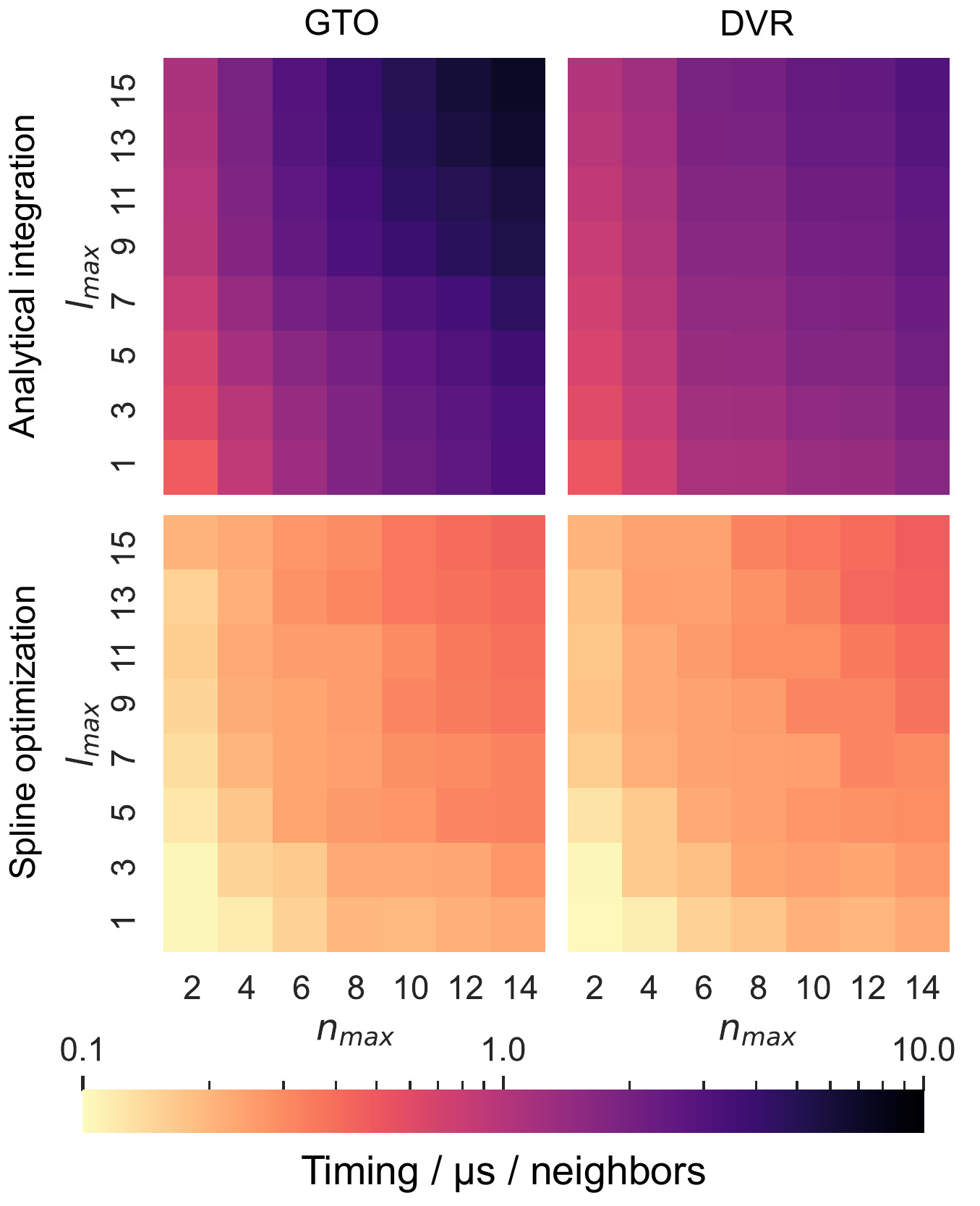}
    \caption{Computational cost for the evaluation of the radial integral and its derivatives with different methods, for structures taken from the QM9 dataset. Note that the dataset has very little influence on this benchmark since the radial integral and its derivative are always evaluated once per neighbor.
    For the splining an accuracy of $10^{-8}$ was chosen.
    }
    \label{fig:ri-qm9}
\end{figure}

\subsection{GTO radial basis}

The Gaussian type orbital radial basis is defined as
\begin{equation} \label{eq:GTO}
\rep<r ||n; \text{GTO}> = \mathcal{N}_{n}r^{n} \exp[-d_{n} r^2],
\end{equation}
where $\facb=1/2\sigma_n^2$, $\sigma_n = r_\text{cut} \max(\sqrt{n},1)/\nmax$, $\mathcal{N}_{n}^2=2/ (\sigma_n^{2n+3} \GA{n+3/2})$ is a normalization factor, and $0\leq n<\nmax$.
In contrast to the displaced Gaussian basis in the original formulation of SOAP,\cite{soap-prb} this choice of radial basis leads to a radial integral that can be evaluated analytically:
\begin{multline}
    \rep<nl||\rij;\text{GTO}> = \pi^{3/2} \exp[-\faca \rij^2] \mathcal{N}_n  \frac{\GA{\frac{n+l+3}{2}}}{\GA{l+\frac{3}{2}}}
    \faca^l \rij^l \\ (\faca+\facb)^{-\frac{n+l+3}{2}} \CHF{\frac{n+l+3}{2}}{l+\frac{3}{2}}{\frac{\faca^2 \rij^2}{\faca+\facb}},
    \label{eq:rad-int-gto-1}
\end{multline}
where ${}_1F_1$ is the confluent hypergeometric function of the first kind. %
Given that the overlap matrix $\mathbf{S}$ between GTOs of the form~\eqref{eq:GTO} can be computed analytically, it is then easy to obtain an orthogonal basis set
\begin{equation}
\rep<n||\text{o-GTO}> = \sum_{n'}{[\mathbf{S}^{-1/2}]}_{nn'}
\rep<n'||\text{GTO}>.
\end{equation}
Thanks to the linear nature of all the operations involved in the evaluation of the density expansion coefficients, the orthogonalization can be applied at any point of the procedure. In the case of the analytical evaluation of Eq.~\eqref{eq:rad-int-gto-1}, it is convenient to first combine the contributions from all the neighbors to the density coefficients \cref{eq:radial-integral-gaussian-density}, and then orthogonalize just once.
In section~\ref{sub:spline}, when computing the coefficients numerically, it is instead more convenient to orthogonalize the radial integral \cref{eq:rad-int-gto-1} directly.

The total time required to compute the radial integral, as well as its derivative with respect to $\rij$ (needed for gradients of the model), is plotted in the top left panel of \cref{fig:ri-qm9} as a function of the expansion parameters $\nmax$ and $\lmax$, and scales roughly linearly with respect to the expansion thresholds (see also the SI for a more detailed figure).
Despite the use of an efficient and robust algorithm which is discussed in \cref{append:1F1}, most of the computational cost in the evaluation of \cref{eq:rad-int-gto-1} is associated with the confluent hypergeometric function ${}_1F_1$.

\subsection{DVR radial basis}

Another possible choice of basis is inspired by the idea of using a numerical, rather than analytical, evaluation of the radial integral. In fact, the numerical integral can be done exactly and with no discretization overhead if we choose the orthonormal DVR radial basis with Gauss-Legendre quadrature rule.~\cite{lightDiscretevariableRepresentations2000}
This basis has the advantage of vanishing at every quadrature point except for one, i.e. $\rep<x||n;\text{DVR}>=\sqrt{\omega_n}\delta(x-x_n)$, which simplifies the numerical radial integral into
\begin{equation}
\rep<nl||\rij;\text{DVR}> = x_n \sqrt{\omega_n} e^{-\faca x_n^2} \mathsf{i}_{l}\left(2\faca x_n \rij \right),
\label{eq:rad-int-dvr}
\end{equation}
where $x_n$ are the quadrature points, distributed across the range $[0,\rcut+3\sigma]$ over which the integrand differs substantially from zero, and the $\omega_n$ are the corresponding quadrature weights.\cite{Abramowitz1972} %
The DVR basis is orthogonal by construction, and only requires evaluating the modified spherical Bessel function rather than the more demanding ${}_1F_1$, leading to a reduction by about a factor of 2 of the cost of evaluating radial integrals (top right panel in \cref{fig:ri-qm9}). Unfortunately, this comes at the price of a less-efficient encoding of structural information, particularly in the limit of sharp atomic Gaussians, as recently shown in Ref.~\citenum{goscinskiRoleFeature2020}.

The computational cost of evaluating the radial integral in the DVR basis is again shown in the upper right-hand panel of Fig.~\ref{fig:ri-qm9}.  The computational cost is reduced by more than half compared to the integral in the GTO basis, although the scaling with the $\lmax$ and especially $\nmax$ parameters remains approximately linear (see plots in the SI).

\subsection{Spline optimization}
\label{sub:spline}

Rather than devising basis functions that allow for a less demanding analytical evaluation of the radial integrals, one can evaluate inexpensively the full radial integral $\rep<nl; r||\denf>$ by pre-computing its value on a grid, and then using a cubic spline interpolator.
For each combination of radial $0\le n < \nmax$ and angular $0\le l \le \lmax $ indices, the integral is tabulated and the spline is computed for the range $[0, r_c]$.  A grid  $\{r_k\}_{k=1}^M$ with constant step size $\Delta$ is used to achieve a constant time complexity for the search of the closest interval $[r_k, r_{k+1}]$ for a distance $r_{ij}\in[r_k, r_{k+1}]$.
Following the implementation of Ref.\citenum{press2007numerical}, the computation of radial terms simplifies to the evaluation of a polynomial of degree 3 in $\rij$ with precomputed coefficients $c_k$ and $d_k$:
\begin{multline}
\rep<nl;\rij||\denf> = \frac{1}{\Delta} \big( (r_{ij}-r_{k+1})c_k+(r_{ij}-r_{k+1})^3 d_k \\ - (r_{ij}-r_{k+1})d_k +
(r_{ij}-r_{k+1})c_{k+1} \\
+(r_{ij}-r_{k+1})^3 d_{k+1} - (r_{ij}-r_{k+1})d_{k+1}  \big).
\label{eq:cubic_spline_eval}
\end{multline}
This expression requires only a small number of multiplications and additions, thus reducing the computational time of the radial integral by avoiding the evaluation of the complex hypergeometric, exponential or Gamma functions present in the analytical GTO and DVR basis sets.
Given that the expression is linear in the coefficients, it is straightforward e.g. to evaluate the coefficients for $\rep<nl||\text{o-GTO}>$ by simply applying the orthogonalization matrix to the coefficients of the $\rep<nl||\text{GTO}>$.
Smooth derivatives ${\partial\rep<nl;\rij||\denf>}/{\partial \rij}$ of this piecewise polynomial function can also be computed by taking the derivative of the polyonmial with minimal additional effort.
As seen in Fig.~\ref{fig:ri-qm9}, splining reduces the computational cost of the radial integrals by almost an order of magnitude, and effectively eliminates the difference between the GTO and DVR basis. %
Thus, the choice of $\rep<x||nl>$ should not be guided by the cost of evaluation, but by a different metric -- for instance, the information efficiency. It has already been shown that GTO encodes linearly regressable information more efficiently than DVR~\cite{goscinskiRoleFeature2020},
implying that the splined GTO basis has a clear advantage overall.
There is evidence that the size of the radial basis set $\nmax$ has a larger influence than the angular expansion threshold $\lmax$ on the accuracy of a SOAP-based potential\cite{roweAccurateTransferable2020}.  Furthermore, a reduction in $\nmax$ redues the cost, not only of the spherical expansion coefficients, but also of evaluating invariants.  Together, these insights all point towards numerical optimization of the radial basis as a promising future line of investigation.

\begin{figure}[bhtp]
    \centering
    \includegraphics[width=\linewidth]{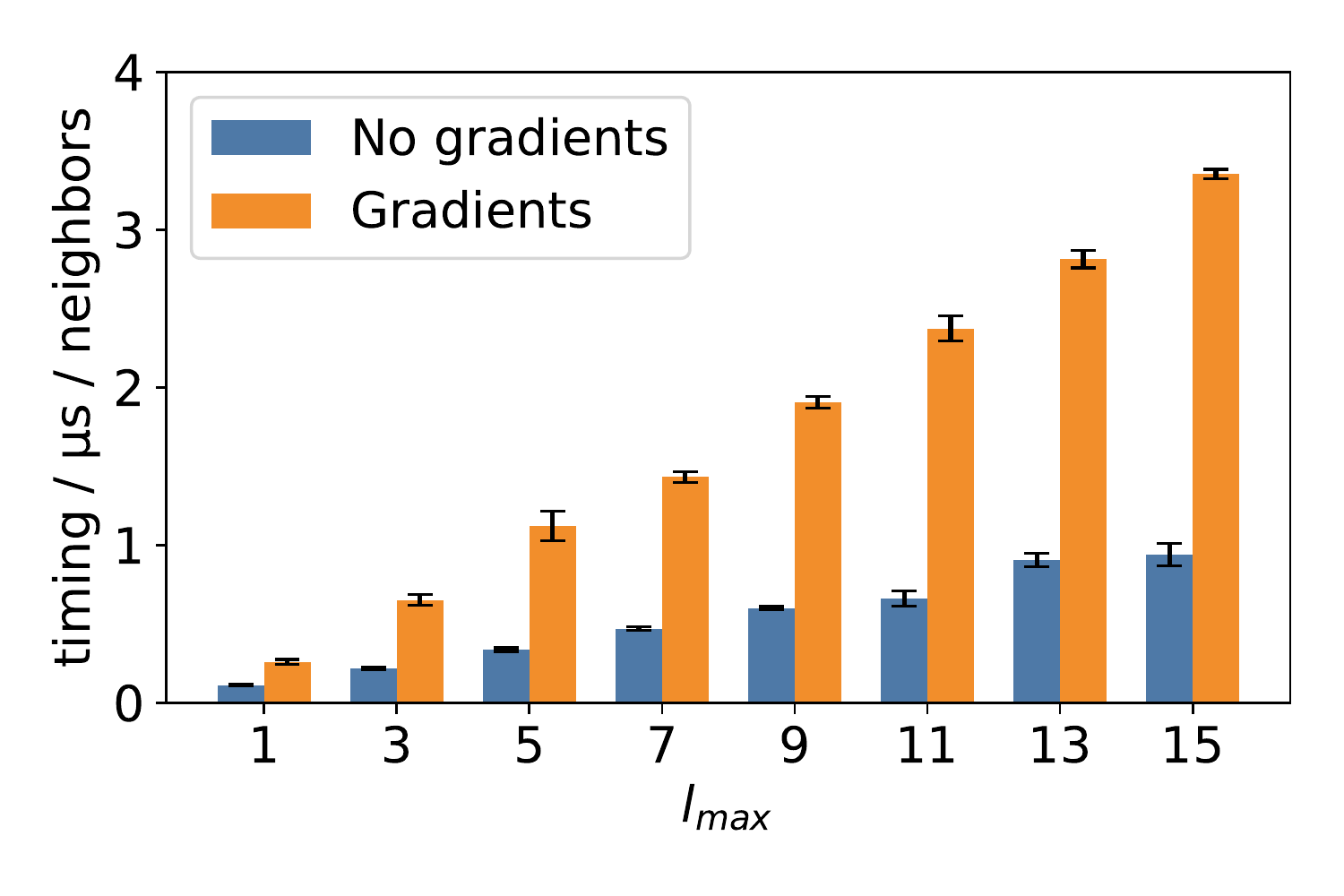}
    \caption{Timings for the computation of the spherical harmonics as a function of the angular expansion threshold for the QM9 dataset.}
    \label{fig:spherical-harmonics-cost}
\end{figure}

\subsection{Spherical Harmonics}
\label{subsec:sph}

In contrast to the relatively obscure special functions needed for the radial integrals, the spherical harmonics needed for the angular part of the density coefficients (cf. Eq.~\eqref{eq:spherical_expansion_coefficients_factorized}) are much more widely used due to their importance in any problem with spherical symmetry.  Correspondingly, there has been much research into finding efficient algorithms to evaluate spherical harmonics, leading to many good algorithms becoming publicly available.  In \rascal{}, we have chosen to implement the algorithm described in \citet{Limpanuparb2014}, which makes use of efficient recurrence relations optimized for speed and numerical stability and is similar to the algorithm implemented in the GNU Scientific Library~\cite{Galassi2009}.  Gradients are computed from analytical expressions. %

As Figure~\ref{fig:spherical-harmonics-cost} shows, the cost scales linearly with the angular expansion parameter $\lmax$, and including gradients consistently increases the cost by a factor of about 4, consistent with the need to compute 3 additional values per spherical harmonic.  The cost to compute the spherical harmonics and gradients is typically comparable to, or larger than, the cost to compute the splined radial integral; this cost is discussed in more detail and in the context of the whole invariants computation in the following section.

\begin{figure*}[htbp]
    \centering
    \includegraphics[width=\linewidth]{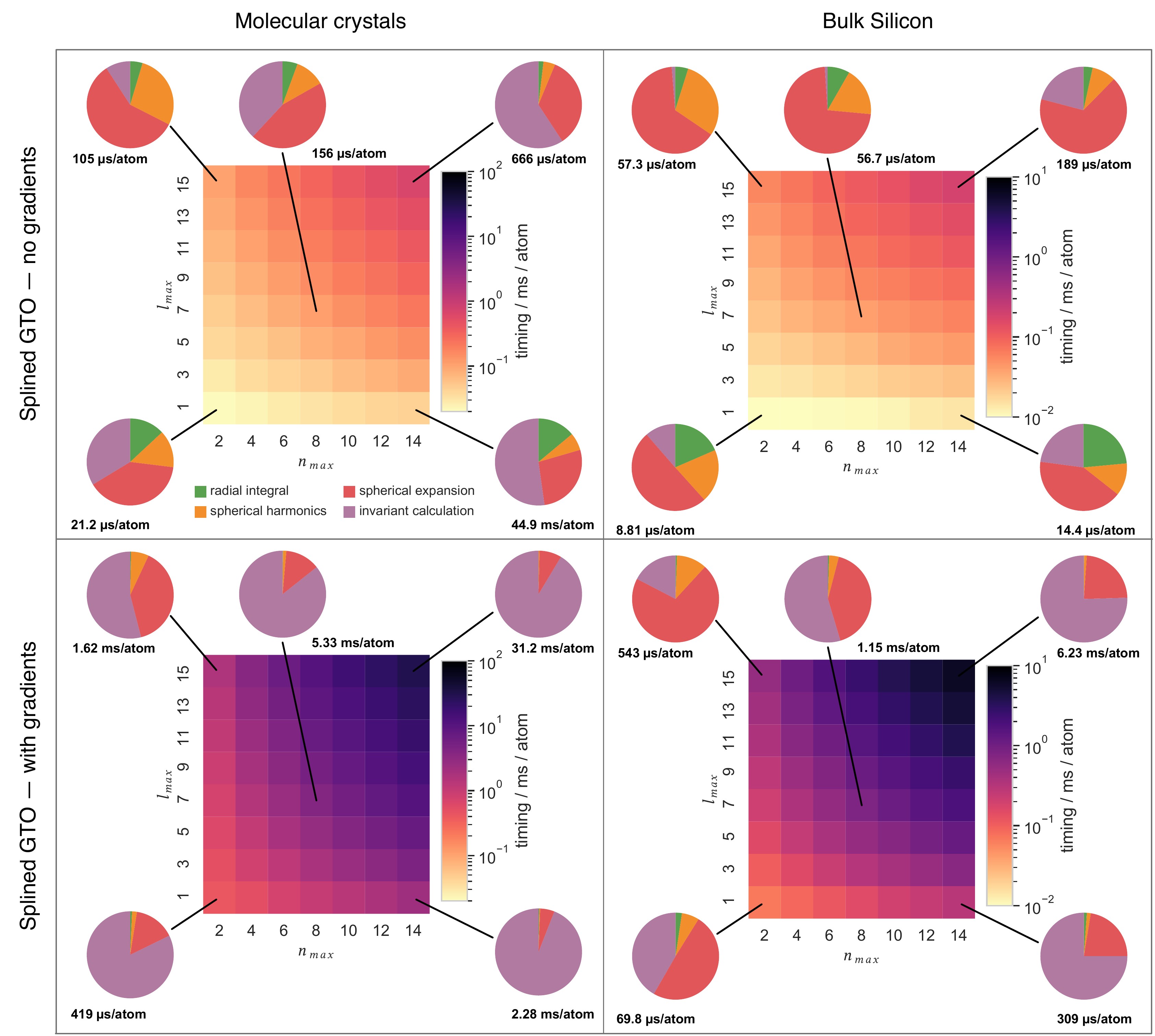}
    \caption{Effect of radial and angular cutoff on overall timing of calculating spherical invariants. (left) molecular crystals dataset (with, on average, 27 neighbors per center, and 4 elements) (right) bulk silicon dataset (16 neighbors per center and a single element). (top) SOAP power spectrum only (bottom) SOAP power spectrum and gradients. All calculations use the GTO radial basis with spline optimization. For selected points, we also show, as pie charts, the relative time spent in the different phases.
    \label{fig:invariant-timings}}
\end{figure*}

\subsection{Spherical Expansion and Invariants}
\label{subsec:se-si}

Having discussed how to implement an efficient procedure to evaluate the radial and the angular terms contributing to the density expansion, let us now consider the cost of the remaining steps to obrain the full SOAP feature vector $\rep<\ennl||\frho_i^2>$.
Figure~\ref{fig:invariant-timings} presents an overview of the timings for all evaluation steps for different $(\nmax,\lmax)$, comparing a dataset of bulk Si configurations and a database of molecular crystals.
For a few selected parameter sets, the figure also shows the breakdown of the evaluation time into the part associated with the evaluation of radial integrals and spherical harmonics for each neighbor, the combination of the two into the full density expansion coefficients, and the calculation of the SOAP invariants.
The spline interpolation makes the cost of radial integrals negligible, and even the evaluation of spherical harmonics usually requires less than 25\%{} of the total timing.  For the silicon dataset, which has only one atomic species, the cost is typically dominated by the combination of radial and angular terms. Indeed, the computational cost of this step scales roughly as $\nneigh  \nmax (\lmax+1)^2$, which can easily dominate the total cost for realistic parameter sets.

For the molecular materials, on the other hand, the evaluation of the invariants becomes more expensive, becoming comparable to the computation the density coefficients.  The difference can be explained as follows.
Given that the coefficients $\rep<\enlm||\rho_i>$ are combined to obtain spherically equivariant representations of the atomic environment by averaging over the group symmetries their tensor products, as outlined in Eq.~\eqref{eq:body-order-x}, their evaluation exhibits a very different scaling.  The cost is independent of the number of neighbors and instead depends strongly on the size of the basis used to expand the atom density as well as on the number of chemical species $\nel$.
For the special case of spherical \emph{invariants} of body order $(\nu + 1) = 3$, corresponding to classic SOAP features\cite{soap-prb}, evaluating Eq.~\eqref{eq:nu2-basis-coupled} essentially amounts to computing an outer product over the $(a,n)$ dimension of expansion coefficients that is then summed over $m$ -- which requires a number of multiplications of the order of $\nel^2\nmax^2 (\lmax+1)^2$.
In summary, the cost of the different steps varies substantially depending on the system, the cutoff, and the expansion parameters, and there is no contribution that dominates consistently in all use cases.

\subsection{Cost of gradients}
\label{sub:gradient-impact}

Evaluating the gradients of the invariant features with respect to the atomic coordinates is a necessary step to compute model derivatives, e.g. forces and stresses for MD simulations -- but it also entails a substantial overhead, as the right-hand panels of Fig.~\ref{fig:invariant-timings} shows.
This overhead is ultimately a consequence of the direct evaluation of the gradients of the features, which requires a separate contraction for each of the $\rep<nn'l|$ components in the SOAP vector,
\begin{multline}
\hspace{-1em}\frac{\partial \rep<nn'l||\frho_i^2>}{\partial \br_j}\!
\propto\! \sum_m \!\!\rep<n'l m||\rho_i>^\star
\frac{\partial \rep<nlm; \br_{ji}||\denf>}{\partial \br_j}  +\ldots
\end{multline}
While some speedup could be attained by reordering the summation, the core issue is the need to compute a separate term for each feature and each neighbor of the central atom, which means that the computational effort, for the typical values of $(\nmax,\lmax)$, is overwhelmingly dominated by the construction of the invariants.  These issues indicate that the evaluation of gradients would benefit from further optimizations -- in particular, trading off modularity for speed by optimizing the expansion coefficients together with the model evaluation. This way, it will be possible to avoid the (re)computation of certain intermediate quantities, analogous to
the optimization of the order of matrix multiplications involved in the evaluation of the chain rule linking the model target and the input atomic coordinates.

\begin{figure}[bhtp]
    \centering
    \includegraphics[width=\linewidth]{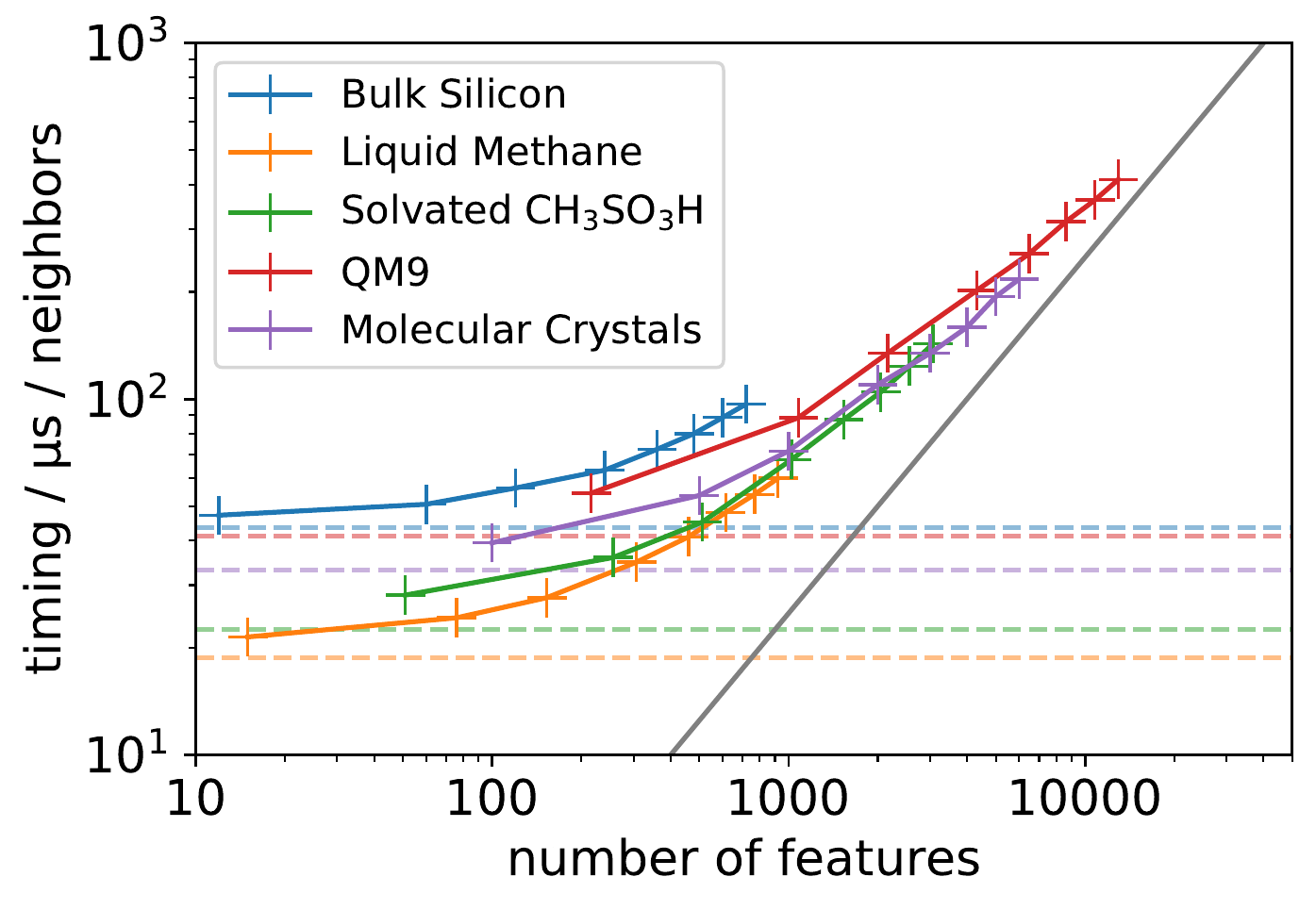}
    \caption{Timing for the calculation of SOAP power spectrum with gradients as function of the requested number of features. Horizontal lines represent the time taken by the spherical expansion step for each dataset. The grey line is a guide for the eye representing a linear relation between time and number of features.
    The ($n_\text{max}$, $l_\text{max}$) parameters used are the following: bulk Silicon (10,11), liquid methane (8,7), solvated CH$_3$SO$_3$H (8,7), QM9 (12,9), and molecular crystals (10,9).}
    \label{fig:si-feat-sparse}
\end{figure}

\subsection{Feature dimensionality reduction}
\label{sub:dim-reduction-sparsification}
A more straightforward, and potentially more impactful, optimization involves performing a data-driven selection to reduce the number of invariant features to be computed and used as inputs of the model.
Even though representations based on systematic orthonormal basis expansions, such as the SOAP power spectrum, provide a complete linear basis to describe 3-body correlations\cite{Glielmo2018,bachmayrAtomicCluster2020}, and even though they do \emph{not} provide an injective representation of an atomic environment\cite{pozdnyakovCompletenessAtomic2020}, one often finds that for realistic structural datasets different entries in the SOAP feature vector exhibit a high degree of correlation. This means that they span a much larger space than what is effectively needed for the prediction of typical atomistic properties.

Therefore, a subset of features -- usually a small fraction of the full set -- can be selected with little impact on the model error.~\cite{imba+18jcp,onat+20jcp}
Both CUR~\cite{maho-drin09pnas} and farthest point sampling (FPS)~\cite{elda+97ieee,ceri+13jctc,imba+18jcp} selection strategies are available in \rascal{}, and can be performed as a preliminary step in the optimization of a model, using Python utility functions.
Feature selection can reduce the time spent both on computing the features, the model parameters and on making predictions (see Section~\ref{sub:kernel-models}). 
For kernel models, and reasonably simple forms of the kernel function, the evaluation of both the features and the kernels scale linearly with $\nfeat$. 
Once a list of selected features has been obtained, their indices $\{q\}\equiv \{(\e_qn_q; \e'_qn'_q;l_q)\}$ can be passed to the \CC{} code.
The sparse feature computation is simply implemented as a selective computation of the pre-selected invariants $\rep<q||\frho_i^2>$. 
The effect of this optimization on the overall cost of computing spherical invariants is shown in Figure~\ref{fig:si-feat-sparse}, with realistic $\nmax$ and $\lmax$ parameters, which are comparable to those used in applications (i.e. $(n_{max},l_{max})$  equal to $(10,12)$ for Si\cite{Bartok2018}, $(8,6)$ for methane\cite{veit+19jctc}, and $(9,9)$ for molecular crystals\cite{musi+19jctc}). 
The overall trend is that of a constant contribution (from the spherical expansion) plus a linear term (from the spherical invariants).  Although most datasets do not reach linear scaling even for the largest number of features,  selecting a small $\nfeat$ can reduce the computational cost by up to an order of magnitude.
The impact of feature dimensionality on both the computational cost and accuracy of models trained on realistic data is discussed in Section~\ref{sec:comparative_benchmarks}; briefly, the features can be sparsified fairly aggressively (up to a factor of about 5--10, depending on the dataset) without any significant impact on the prediction error.

\section{Comparative Benchmarks}
\label{sec:comparative_benchmarks}

Now that we have analyzed separately the different components of the calculation of the SOAP features, we turn our attention to the end-to-end benchmarking of a full energy and force evaluation, similar to what one would encounter when running a MD simulation.
As in the previous Section, we run comprehensive tests on each of the five datasets described in Section~\ref{sub:methods-datasets}, and we report here those that are most telling of the scaling of the different terms with the system parameters, most notably the neighbor density and the number of chemical elements. We include a simple but complete implementation of kernel ridge regression, a framework that is often used together with SOAP features and that allows us to comment on the interplay between the calculation of the representation and the model.
Thus, we can compare the computational effort associated with the use of \rascal{} with that of QUIP, an existing, well-established code to train and evaluate Gaussian approximation potentials (GAPs) based on SOAP features, and investigate the effect of the various optimizations described above on the overall model efficiency.

\subsection{Existing implementations}
\label{sub:existing-codes}

Over the past couple of years, several codes have been released that can be used to fit and run ML potentials supporting different representations, especially for neural-network type potentials such as n2p2~\cite{N2P2} (which uses Behler-Parrinello ACSF~\cite{Behler2011}), ANI-1~\cite{Smith2017}, PANNA~\cite{lotPANNAProperties2020}, or DeepMD\cite{wang+18cpc}.
Here we focus on kernel methods, for which there is a smaller number of actively used codes.  The first, and still widely adopted, is the QUIP library, part of the libAtoms framework~\cite{libAtoms}, which has been used for almost all published Gaussian approximation potentials (GAPs)~\cite{bart+10prl,deri-csan17prb,veit+19jctc,Mocanu2018,Caro2018,Caro2019,zhangPartitioningSulfur2020} and continues to be actively maintained.  Other kernel-learning potential packages of note are GDML, which implements the ``gradient-domain machine learning'' method of \citet{Chmiela2017} (the full-kernel equivalent of the sparse kernel model we implement here), and QML~\cite{Christensen2017}, which notably implements the FHCL-type representations~\cite{Faber2018} and the OQML framework\cite{Christensen2019}.  We finally note for completeness several codes used for linear high-body-order models, such as the SNAP method~\cite{thompsonSpectralNeighbor2015} implemented in LAMMPS~\cite{plimpton1995fast}, aPIPs~\cite{vanderoordRegularisedAtomic2020} and ACE~\cite{drau19prb, bachmayrAtomicCluster2020} implemented in JuLIP~\cite{JULIP}, and the NICE descriptors~\cite{nigamRecursiveEvaluation2020} implemented in a separate code~\cite{NICE-REPO} interfaced with \rascal{} (see Section~\ref{sec:experimental_features}).
Here we focus only on the QUIP code, which is the most mature implementation available and matches most closely the application domain of \rascal{}. %

\subsection{Kernel models}
\label{sub:kernel-models}

To benchmark the performances of \rascal{} in the context of the GAP framework typically used to build potentials with SOAP, we implemented the same regression scheme used in QUIP to build a MLIP based on the SOAP power spectrum representation. We summarize the key ideas, emphasizing the aspects that are important to achieve optimal performance.
In a GAP, as in the vast majority of regression models based on atom-centred features, the energy is defined as a sum of atomic contributions
\begin{equation}\label{eq:additive-energy-model}
E(A) \equiv \rep<E||A> = \sum_{i\in A} E(A_i)\equiv\sum_{i\in A} \rep<E||A_i>
\end{equation}
where $A_i$ indicates a local environment centered on atom $i$.
An accurate, yet simple and efficient GAP can be built using a ``projected process approximation''\cite{snelson2006sparse} form of kernel ridge regression, that  mitigates %
the unfavorable scaling with train set size $\ntrain$ of the cost of fitting (cubic) and predicting (linear) energies using a ``full'' ridge regression model.
A small, representative subset $M$ of the atomic environments usually found in the training set -- the so-called ``active'', ``pseudo-'' or ``sparse'' points -- is used, together with a positive-definite kernel function $k$, as a basis to expand the atomic energy
\begin{equation}\label{eq:sparse-gp}
E(A_i)  = \sum_{I\in M}\delta_{\e_i \e_I}\rbraket{E; \e_I}{M_I} k(M_I,A_i),
\end{equation}
where $M_I$ indicates the $I$-th sparse point, $\rbraket{E; \e_I}{M_I}$ indicates the regression weights, and a separate energy model is determined for each atomic specie, which also means that the active set is partitioned with respect to the central atom type.
The sparse model~\eqref{eq:sparse-gp} exhibits a much more favorable scaling with training set size, both during fitting ($\mathcal{O}(\ntrain\nactive^2+\nactive^3)$, for the implementation in \rascal{}) 
and when predicting a new structure ($\mathcal{O}(\nactive)$). %
Obviously, the accuracy of the approximation relies on a degree of redundancy being present in the training set, and in practice a suitable size of the active set $M$ scales with the ``diversity'' of the training set. Usually, however, an accuracy close to that of a full model can be reached even with $\nactive\ll \ntrain$.
The gradient of the energy with respect to the coordinates of an atom $j$ can be obtained as a special case of the general form~\eqref{eq:energy-derivative}
\begin{equation}
\grad_j \rbraket{E}{A} = \sum_{I\in M} \rbraket{E; \e_I}{M_I} \sum_{i\in A} \delta_{\e_i \e_I}  \grad_{j}  k(M_I,A_i) ,  \label{eq:gpr-force}
\end{equation}
and the virial (the derivative with respect to deformations $\bm{\eta}$ of the periodic cell)
as a special case of~\eqref{eq:energy-virial}
\begin{multline}
\pdv{}{\bm{\eta}}\rbraket{E}{A} = \sum_{I\in M} \rbraket{E; \e_I}{M_I} \sum_{i\in A} \delta_{\e_i \e_I} \sum_{j\in A_i} \\   \br_{ji} \otimes \grad_j  k(M_I,A_i).
     \label{eq:gpr-stress}
\end{multline}
In both Eqs.~\eqref{eq:gpr-force} and~\eqref{eq:gpr-stress}, the sum over the neighbors of atom $i$ extends also over periodic replicas of the system.
Both equations require the evaluation of kernel gradients, that can in turn be expressed using the chain rule in terms of the derivatives of the kernel function with respect to atomic features, and the atomic gradient of such features:
\begin{multline}
\grad_j  k(M_I,A_i) = \sum_{q} \grad_{j} \rbraket{q}{A_i;\text{rep}} \pdv{k(M_I,A_i)}{\rep<q||A_i>}.
\end{multline}
When computing the model derivatives it is important to contract the sums in the optimal order, by first summing the derivatives of the kernel over the active set. For instance, for the force,
\begin{multline}
\grad_j \rbraket{E}{A} =  \sum_{i\in A} \delta_{\e_i \e_I} \sum_q  \grad_{j}\rbraket{q}{A_i;\text{rep}} \\
\times \left[\sum_{I\in M} \rbraket{E; \e_I}{M_I}  \pdv{k(M_I,A_i)}{\rep<q||A_i>}\right].
\label{eq:gpr-force-order}
\end{multline}
This form shows that the cost of evaluating forces scales with $\nfeat \nneigh \nactive$, indicating how the reduction of the number of sparse points \emph{and} features combine to accelerate the evaluation of energy and forces using a sparse GAP model.

The fitting procedure that is implemented in \rascal{} has been discussed in Ref.~\cite{csan+20book}, and we do not repeat it here. It only requires the evaluation of kernels and kernel derivatives between the active set environments, and the environments in the structures that are part of the training set, and is usually limited by memory more than by computational expense.
In the benchmarks we present here we adopt the polynomial kernel which has been widely used to introduce non-linearity into SOAP-based GAP models~\cite{bart+10prl,deri-csan17prb,veit+19jctc,Mocanu2018,Caro2018}:
\begin{equation}
k_\zeta(M_I,A_i) = \left[\sum_q \rbraket{M_I; \text{rep}}{q} \rbraket{q}{A_i; \text{rep}}\right]^\zeta,
\end{equation}
whose derivative can be simply computed as
\begin{equation}
\pdv{k_\zeta(M_I,A_i)}{\rep<q||A_i>} = \zeta \rbraket{M_I; \text{rep}}{q} k_{\zeta-1}(M_I,A_i)  .
\end{equation}

\subsection{Benchmarks of sparse models}
\label{sub:benchmarks-sparsification}

\begin{figure*}[bhtp]
    \centering
    \includegraphics[width=\textwidth]{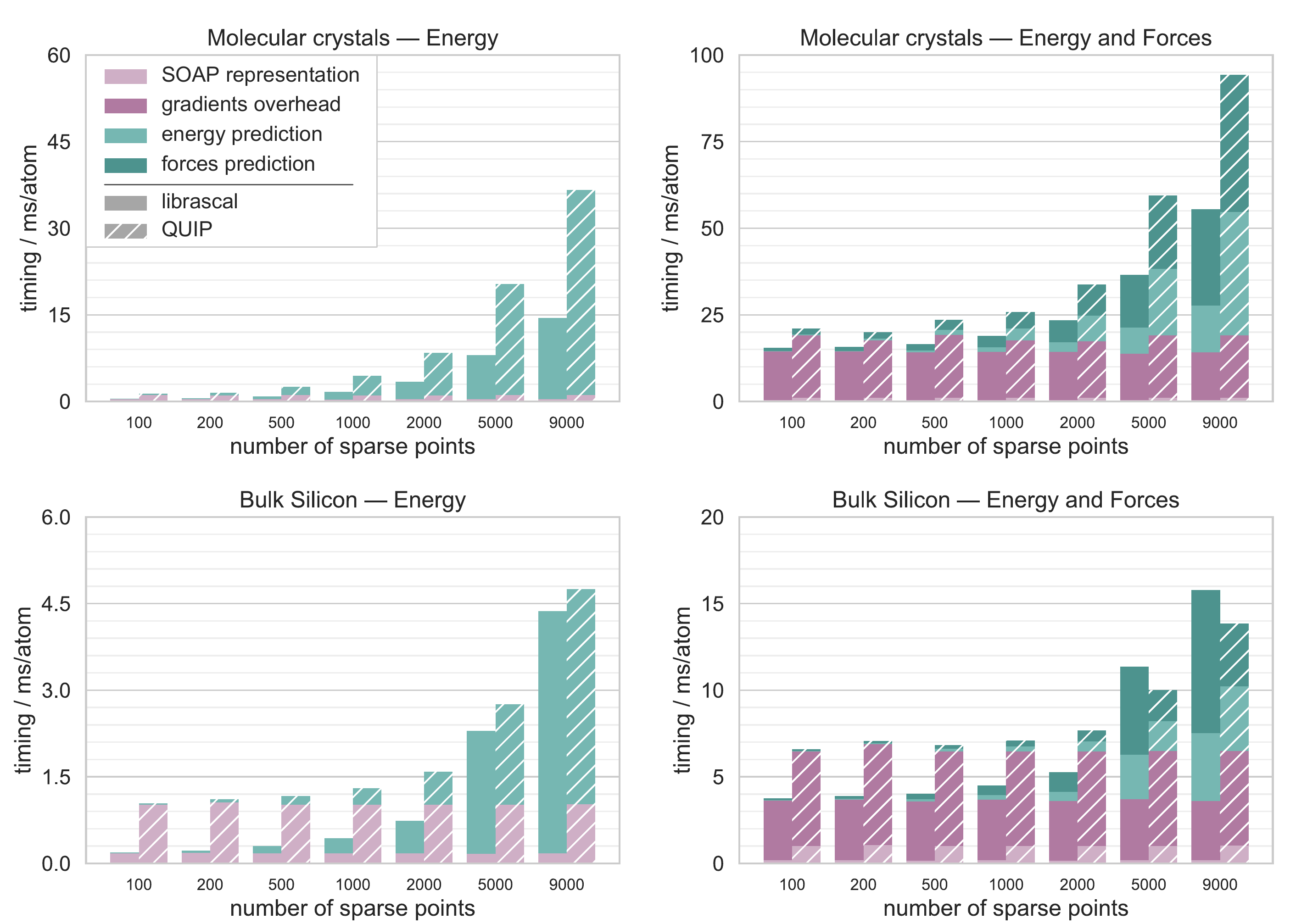}
    \caption{
        Prediction timings for GAP models as a function of the number of sparse points, with (right) and without (left) the evaluation of forces, with minimal feature sparsification, i.e., just enough to eliminate redundant symmetric terms (these are retained in \rascal{} for simpler bookeeping). We used all unique SOAP features for each system in this figure, meaning 6660 features for the molecular crystals and 715 features for bulk silicon.
    }
    \label{fig:soap-model-bench-light-feat-sparse}
\end{figure*}

\begin{figure*}[htbp]
    \centering
    \includegraphics[width=0.9\textwidth]{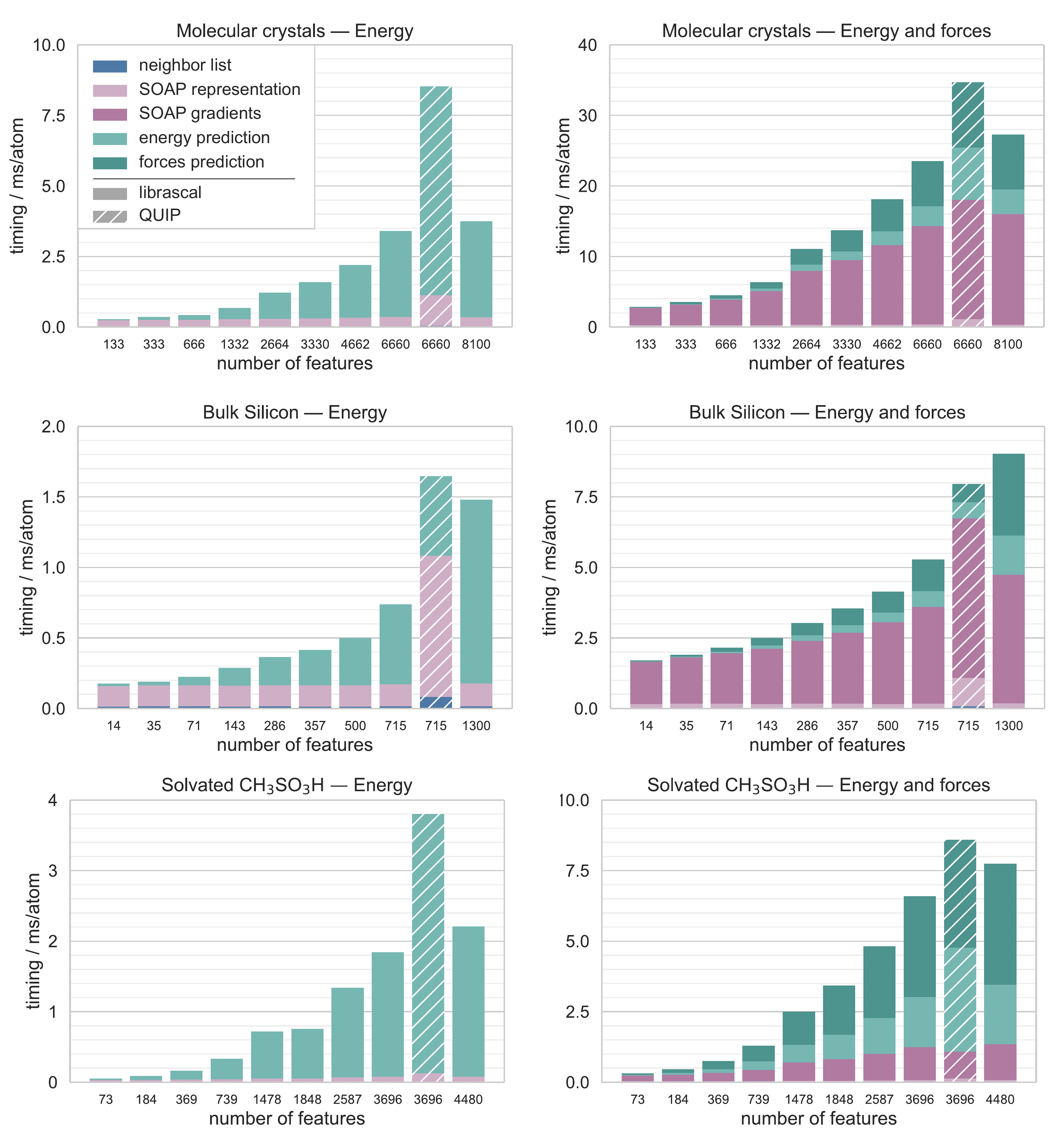}
    \caption{Prediction timings for GAP models as a function of the number of features, with (right) and without (left) the evaluation of forces.  All models use 2000 sparse points for the sparse kernel basis.  The rightmost column in each plot shows the cost with some redundant features, which are computed by default in \rascal{} for simpler bookkeeping. In practical applications, though, we recommend these be eliminated automatically through feature sparsification.}
    \label{fig:features_sparsification}
\end{figure*}

Having summarized the practical implentation of a sparse GPR model based on SOAP features, we can systematically investigate the effect of the sparsification parameters -- number of sparse environments $\nactive$ and number of sparse features $\nfeat$ -- on the different components of an energy and force calculation.
Figures~\ref{fig:soap-model-bench-light-feat-sparse} and~\ref{fig:features_sparsification}  %
show the full cost of evaluating a MLIP for different classes of materials, both with and without the evaluation of forces, for different levels of sparsification in terms of both $\nactive$ and $\nfeat$.
The cost is broken down in the contributions from the evaluation of the the neighbour list, the representation, and model evaluation (prediction) steps.

Figure~\ref{fig:soap-model-bench-light-feat-sparse} shows that, when using the full feature vector in the model\footnote{In practice, to match the number of features computed by QUIP, we use a mild feature sparsification in \rascal{} that corresponds to the same use of the $\rep<\e n_1; \e n_2; l| = \rep<\e n_2; \e n_1; l|$  symmetry that is implemented in QUIP.}, the evaluation of the kernels contributes substantially to the cost of predicting energies.
In QUIP this cost, which scales linearly with the number of active points, matches the cost of evaluating the representations -- independent of $\nactive$, since the same number of representations must always be computed for the target structure, at $\nactive\approx 5000$ for Si, and  $\nactive\approx 500$ for the molecular crystals.
Due to the optimization of the feature evaluation step, in \rascal{} the kernel evaluation dominates down to even smaller $\nactive$. Note also the lower cost of the kernel evaluation for the molecular crystals in \rascal{}, which can be explained by the fact that only the features associated with chemical species that are present in each structure are computed, while in QUIP they yield blocks of zeros that are multiplied to compute scalar products.
Evaluating also forces (right-hand panels of Fig.~\ref{fig:soap-model-bench-light-feat-sparse}) introduces a very large overhead to feature calculation (up to one order of magnitude, as discussed in the previous Section) and roughly doubles the cost of  model prediction. Since the cost of feature evaluation is independent on $\nactive$, the active set can be expanded up to thousands of environments before the model evaluation become comparable to feature evaluation.

In order to accelerate calculations further, it is then necessary to reduce not only the time needed to compute the model, but also the time needed to compute the representation itself.
In Fig.~\ref{fig:si-feat-sparse} we showed how restricting the evaluation of SOAP features to a smaller subset of the $\rep<\ennl|$ indices reduces by up to an order of magnitude the cost of evaluating the feature vector and its gradients.
Figure~\ref{fig:features_sparsification} demonstrates how this speedup combines with the acceleration of the model evaluation step, whose nominal complexity also scales linearly with $\nfeat$, for a intermediate size of the active set $\nactive=2000$.
For simple, single-component systems such as bulk silicon the cost saturates to that of evaluating the density expansion coefficients, and so the overall speedup that can be achieved by feature sparsification is limited to about a factor of two or three with respect to the full SOAP power spectrum. For multi-component systems, such as the CSD dataset or the solvated \ce{CH3SO3H} dataset, a speedup of nearly an order of magnitude is possible.

\begin{figure*}[hbtp]
    \centering
    \includegraphics[width=1.0\linewidth]{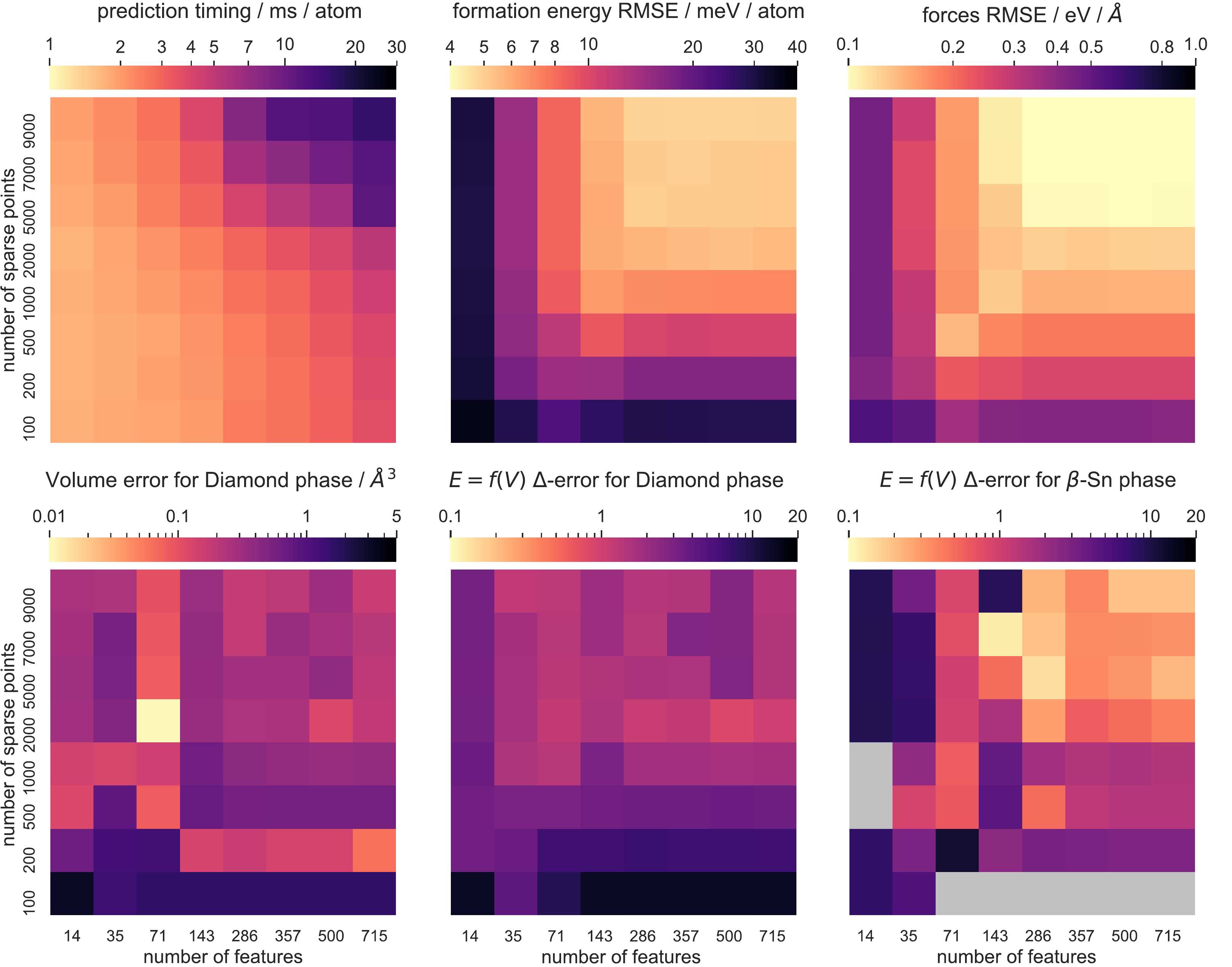}
    \caption{Evaluation of the GAP model performance for bulk Silicon.
    We present the evaluation cost and corresponding error as a function of the number of sparse training point and features selected. From left to right, top to bottom: time required to evaluate the model, root mean square error for the predicted energies and forces, absolute error in the predicted volume compared in the Diamond phase, and $\Delta$-error --- see equation~\eqref{eq:delta-measure} --- for the energy/volume curve for Diamond and $\beta$-Sn phases. For all errors, the reference are the values from DFT calculations~\cite{Bartok2018}.
    \label{fig:accuracy-cost-si}}
\end{figure*}

\subsection{Accuracy-cost tradeoff}

While the performance optimization discussed in Section~\ref{sec:impl-bench} can dramatically increase the efficiency of a MLIP based on SOAP features and sparse GPR, one should obviously ensure that models with reduced $\nactive$ and $\nfeat$ still achieve the desired accuracy.
The data-driven determination of the most representative and diverse set of features and samples is a very active area of research, using both unsupervised\cite{elda+97ieee,maho-drin09pnas,bart-csan15ijqc,imba+18jcp,onat+20jcp} and, very recently, semi-supervised\cite{cersonsky2020arxiv} criteria to select an optimal subset.
Here we use the well-established  FPS to sort features and environment in decreasing order of importance, starting from the full list of environments for the Si dataset and a pool of 715 features, corresponding to $\nmax=10, \lmax=12$. 
We train MLIPs to reproduce energy and forces and report the four-fold cross-validation error as well as the cost for evaluating the energy and its gradients in Figure~\ref{fig:accuracy-cost-si}, using only the ``best'' $\nactive$ active points and $\nfeat$ features.
We also report the ``$\Delta$'' measure introduced in Ref.~\citenum{leja+16science} as an indication of the ability of the ML model to reproduce properties that are indirectly related to the accuracy of the PES~\cite{szla+14prb,Bartok2018}:
\begin{equation}
\Delta = \sqrt{\frac{\int_{0.95\,V_0}^{1.06\,V_0} \left[E^\text{GAP}(V) - E^\text{DFT}(V)\right]^2 \text{d}V}{0.12 \ V_0}}
\label{eq:delta-measure}
\end{equation}
where $E^\text{GAP}$ and $E^\text{DFT}$ are the GAP and DFT energies relative to the diamond energy minimum, and $V_0$ is the volume of the minimum DFT energy structure for each phase.

The results clearly indicate that it is possible to considerably reduce the cost of the MLIP with little impact on the accuracy of the model.  Severe degradation of model performance occurs in the regime in which the computational cost is dominated by the calculation of the density expansion coefficients, suggesting that further optimization of the evaluation of $\rep<\enlm||\rho_i>$ might not be exceedingly beneficial to most practical use cases.

\todorev{\MC{would be nice to have a similar figure for another dataset, say the molecular crystals or CH3SO3H}}

\section{Experimental features}
\label{sec:experimental_features}

The spherical expansion coefficients can also be used to compute equivariant features and kernels\cite{glielmoAccurateInteratomic2017,gris+18prl}, as well as higher-body-order invariants\cite{Willatt2019,drau19prb}.
This evaluation is easily and efficiently done with an external library, as it is done in the current implementation\cite{NICE-REPO} of the N-body iterative contraction of equivariants (NICE) framework\cite{nigamRecursiveEvaluation2020}.
Furthermore, \rascal{} contains experimental implementations of other representations based on the SOAP framework, for example the bispectrum\cite{soap-prb,thompsonSpectralNeighbor2015} and the $\nu=2$ equivariants that underlie the $\lambda$-SOAP kernels\cite{gris+18prl} (which is also available as an independent implementation\cite{TENSOAP}).
As development progresses, these libraries will be further integrated with \rascal{}, harmonizing and streamlining the user-facing APIs, and achieving the best balance between modularity and evaluation efficiency.

\todorev{CONTRACTED EQUIVARIANTS NEW TRICK}

\section{Conclusions}
\label{sec:conclusion}

In this paper we have made practical use of recent insights into the relationships between several families of representations that are typically applied to the construction of machine-learning models of the atomic-scale properties of molecules and materials. We have demonstrated how these insights can be translated into algorithms for more efficient computation of these representations, most notably SOAP, but also the atom-density bispectrum and the $\lambda$-SOAP equivariants.
We have shown how the radial basis used to expand the density can be chosen at will and computed quickly using a spline approximation. Together with a fast gradient evaluation, this reduces the cost of computing the density expansion to the point where it is no longer the rate limiting step of the calculation in typical settings.
Further optimizations can be obtained by a ``lossy'' strategy, which trades off some accuracy for efficiency by discarding redundant or highly correlated entries in both the active set of a projected-process regression model and in the invariant features.
We have implemented all these optimizations in \rascal{}, a modular, user-friendly and efficient open-source library purpose-built for the computation of atom-density features (especially SOAP).

In order to test these optimizations in practice, we have run benchmarks over different kinds of datasets spanning elemental materials as well as organic molecules in isolation, in crystalline phases, and in bulk liquid phases. 
Using one of the most widespread codes for the training and evaluation of SOAP-based machine-learning interatomic potentials as a reference, we have found that our implementation of the SOAP representation is much faster, but that the advantage is less dramatic when considering also the calculation of a kernel model, which scales with the number of features, and that of the gradients, which is dominated by a term that scales with the number of neighbors in both codes. 
Feature selection, however, addresses both these additional overheads, and allows for an acceleration of the end-to-end evaluation time of energy and forces by a factor anywhere between four and ten with minimal increase in the prediction errors.
Our tests show that in the current implementation, when using realistic values of the parameters, the different steps of the calculation contribute similarly to the total cost, indicating that there is no single obvious bottleneck. Further improvements, although possible, should consider the model as a whole and especially improve the accuracy/cost balance of lossy model compression techniques.

\appendix

\section{Efficient implementation of ${}_1F_1$}
\label{append:1F1}
The confluent hypergeometric function of the first kind is defined as
\begin{equation}
    \CHF{a}{b}{z}= \sum_{s=0}^{\infty} \frac{(a)_s}{(b)_s s!} z^{s},
\end{equation}
where $(a)_s$ is a Pochhammer’s symbol (Ref.~\citenum{dlmf}, Chap. 5.2(iii)).
To efficiently compute \cref{eq:rad-int-gto-1}, we implement a restricted version of ${}_1F_1$
\begin{equation}
G(a,b,\rij) = \frac{\Gamma(a)}{\Gamma(b)}  \exp[-\faca \rij^2] \CHF{a}{b}{\frac{\faca^2 \rij^2}{\faca+\facb}},
\end{equation}
where $a=\frac{n+l+3}{2}$ and $b=l+\frac{3}{2}$. We take into account that the arguments of ${}_1F_1$ are real and positive and we avoid its artificial overflow by using the asymptotic expansion (Eq.~13.2.4	and Eq.~13.7.1 in Ref.~\citenum{dlmf})
\begin{equation}
    \underset{z\rightarrow\infty}{\lim} \CHF{a}{b}{z} = e^z z^{a-b} \frac{\Gamma(b)}{\Gamma(a)} \sum_{s=0}^{\infty} \frac{(b-a)_s(1-a)_s}{s!} z^{-s},
\end{equation}
since the exponential in Eq.~\eqref{eq:rad-int-gto-1} can be factorized $\exp[\frac{\faca^2 \rij^2}{\faca+\facb}]\exp[-\faca \rij^2]=\exp[\faca \rij^2 (\frac{\faca}{\faca+\facb}-1)]$ and $\frac{\faca}{\faca+\facb}-1 < 0$.
Note that $G$ is implemented as a class so that the switching point between the direct series and the asymptotic expansion evaluations is determined at construction for particular values of $a$ and $b$ using the bisection method.

For each value of $n$, the function $G$ and its derivatives with respect to $\rij$ can be efficiently evaluated using the two step recurrence downward relation
\begin{align}
    G(a+1,b+1,\rij) =& \frac{\faca^2 \rij^2}{\faca+\facb} G(a+2,b+3,\rij) \nonumber \\
    &+ (b+1) G(a+1,b+2,\rij), \\
    G(a,b,\rij) =& \frac{\faca^2 \rij^2}{\faca+\facb} \frac{a-b}{a} G(a+1,b+2,\rij) \nonumber \\
    &+ \frac{b}{a} G(a+1,b+1,\rij),
    \label{eq:1f1-recurrence}
\end{align}
with $\pdv*{G(a,b,\rij)}{\rij} = \frac{2\faca^2 \rij}{\faca+\facb}G(a+1,b+1,\rij) - 2\faca \rij G(a,b,\rij)$.
We found empirically that only the downward recurrence relation was numerically stable for our range of parameters.
Note that $a+1$ corresponds effectively to steps of $l+2$ so computing $G$ and $\dv{G}{\rij}$ for all $l\in[0,\lmax]$ and all $n\in[0,\nmax[$ requires $4\nmax$ evaluations when using this recurrence relation.

\section{Derivatives of the energy function}
\newcommand{\bh}[0]{\mbf{h}}
We have defined an atom centered energy model such that the energy associated with structure $A$ can be written as in Eq.~\eqref{eq:additive-energy-model}, $E(A)= \sum_{i\in A} E(A_i)$.
The structure $A$ is determined by the set of atomic coordinates and species $\{\br_i,\e_i\}$ and (for periodic structures) unit cell vectors $\{\bh_1,\bh_2,\bh_3\}$. The atom-centred environment $A_i$ is entirely characterized by the atom centered vectors $\{\br_{ji}=\br_j-\br_i\}$ with $r_{ji}<\rcut$.
The derivatives of $E$ with respect to the position of atom $k$ (the negative of the force acting on the atom) can be computed using the chain rule
\begin{equation}
\pdv{E(A)}{\br_k} = \sum_{i\in A} \pdv{E(A_i)}{\br_k} =  \sum_{i\in A} \sum_{j\in A_i} \pdv{E(A_i)}{\br_{ji}} \cdot \frac{\partial \br_{ji}}{\partial \br_{k}}.\label{eq:energy-derivative}
\end{equation}
Here index $j$ runs over the neighbors of atom $i$, which include periodic images, if the system is periodic.
The term ${\partial \br_{ji}}/{\partial \br_{k}}$ is zero unless $k=i$ (in which case it evaluates to $-1$) or if $j=k$ (in which case it evaluates to $1$). In the periodic case, the derivative with respect to $\br_k$ has to be interpreted as one in which all periodic images of atom $k$ are displaced simultaneously. Thus, when the neighbor $j$ is a periodic image of $k$, ${\partial \br_{ji}}/{\partial \br_{k}}=1$, and if $k=i$ the total contribution of the periodic images of $i$ is zero.

For the virial, we need to compute the derivative of the energy with respect to infinitesimal strain deformations of the unit cell $\bm{\eta}$.
Using the atom-centred decomposition, and applying the chain rule as above one gets
\begin{multline}
\pdv{E(A)}{\bm{\eta}} = \sum_{i\in A} \pdv{E(A_i)}{\bm{\eta}} = \sum_{i\in A} \sum_{j\in A_i}  \pdv{E(A_i)}{\br_{ji}} \frac{\partial \br_{ji}}{\partial \bm{\eta}}\\=
\sum_{i\in A} \sum_{j\in A_i} \pdv{E(A_i)}{\br_{ji}} \otimes \br_{ji}\label{eq:energy-virial}
\end{multline}
where again $j$ runs over all the neighbors including periodic images. It is interesting to note that the periodic images of $i$ will have a non-zero contribution to the virial.

\section*{Supplementary materials}

The supplementary materials contain figures describing the benchmark results for all of the datasets mentioned in this work.

\section*{Data availability}
Data supporting the findings in this paper are available from public repositories as referenced, or upon reasonable request to the authors.
The source code of \rascal{} is available from an open-source repository\cite{LIBRASCAL}, and workflows that can reproduce the benchmarks reported in this paper are available from a separate repository\cite{rascal_benchmarks}.

\begin{acknowledgments}
FM, MV, MS and MC support by the NCCR MARVEL, funded by the Swiss National Science Foundation (SNSF).
AG and MC acknowledge support from the Swiss National Science Foundation (Project No. 200021-182057).
GF acknowledges support by the European Center of Excellence MaX, Materials at the Exascale - GA No. 676598.

\end{acknowledgments}

\onecolumngrid
\clearpage
\newpage

\end{document}